\documentclass[letterpaper, 10 pt, conference]{ieeeconf}  

\IEEEoverridecommandlockouts                              

\usepackage{graphicx}
\usepackage{hyperref}
\usepackage{makecell}
\usepackage{soul} 
\usepackage{xcolor} 
\usepackage{booktabs} 

\title{\LARGE \bf
	Spatial Insight: How Data-Driven Regions of Interest Selection Enhances Single-Trial P300 Classification in EEG-Based BCIs*
}

\author{Eva Guttmann-Flury$^{1}$, Jian Zhao$^{2}$, and Mohamad Sawan$^{3}$
	\thanks{*This work is supported in part by STI 2030
		Major Projects, the National Key Research and Development Program of China (Grant No. 2022ZD0208500).}
	\thanks{$^{1}$Eva Guttmann-Flury is with the Department of Micro-Nano Electronics and the MoE Key Laboratory of Artificial Intelligence,
		Shanghai Jiao Tong University, Shanghai, China, and a visiting Postdoctoral Fellow in CenBRAIN Neurotech, School of Engineering, Westlake University, Hangzhou, China
		{\tt\small eva.guttmann.flury@gmail.com}}%
	\thanks{$^{2}$Jian Zhao is with the Department of Micro-Nano Electronics and the MoE Key Laboratory of Artificial Intelligence,
		Shanghai Jiao Tong University, Shanghai, China
		{\tt\small zhaojianycc@sjtu.edu.cn}}%
	\thanks{$^{2}$Mohamad Sawan is with the CenBRAIN Neurotech, School of Engineering,  
		Westlake University, Hangzhou, China
		{\tt\small sawan@westlake.edu.cn}}%
}

\begin{document}

	\maketitle
	\thispagestyle{empty}
	\pagestyle{empty}

	\begin{abstract}
		
		EEG-based Brain-Computer Interfaces (BCIs) frequently face spatial specificity limitations in detecting single-trial P300 potentials, a neurophysiological hallmark leveraged for both BCI control and neurodegenerative disease diagnostics. We present a novel framework combining eLORETA source localization with cross-subject functional connectivity to identify stable regions of interest (ROIs) across sessions. Analyzing 62-channel EEG data from 31 subjects (63 sessions, 2,520 trials), we demonstrate that phase-lagged connectivity metrics can reliably isolate task-relevant hubs in deeper cortical-subcortical structures like the insula and parietal regions — critical for Alzheimer’s disease biomarkers. By integrating spatially stable ROIs with dynamic temporal agreement, our hybrid classification systematically outperforms whole-brain approaches in different frequency bands (up to 5.4\% depending on the connectivity method and the spectral range) while maintaining millisecond-level temporal precision.
		
		To the best of our knowledge, this is the first study to establish cross-subject ROI consensus through source-space connectivity, bypassing scalp EEG’s depth constraints to probe Alzheimer’s-relevant networks. The framework's robustness to noise and compatibility with portable systems offer significant potential for global deployment in early neurodegenerative disease detection. Future integration of individualized anatomical data or adaptive parameter optimization could refine this tool for clinical deployment, enhancing the current max accuracy of 81.57\% in the 1-15 Hz range.
	\end{abstract}

	\section{INTRODUCTION}
	
	Brain-Computer Interfaces (BCIs) translate neural activity into actionable commands for external devices, bypassing traditional neuromuscular pathways. Non-invasive BCIs using electroencephalography (EEG) offer millisecond-level temporal resolution and portability but face limitations due to low spatial resolution and susceptibility to noise and artifacts. These issues stem from the complex biophysics of electrical signal propagation through heterogeneous head tissues, leading to an ill-posed inverse problem with infinite mathematically valid source configurations \cite{Marino_2024}.
	
	Among EEG-based BCI systems, P300 paradigms are widely adopted for their robustness and versatility, leveraging the brain’s stereotyped response to rare or task-relevant stimuli. Applications span from assistive communication (e.g., for locked-in patients) to neurodiagnostics (e.g., Alzheimer’s biomarkers) and even healthy-user domains \cite{Kalra_2023}.

	
	The P300 response, a neurophysiological hallmark of cognitive processing, is characterized by robust spatiotemporal properties: a biphasic waveform with frontocentral P3a (150 -- 280 ms post-stimulus) and parietal P3b (300 -- 500 ms) components \cite{Hedges_2016, Takano_2014}. This potential originates from distributed cortical-subcortical generators \cite{Sabeti_2016, Citherlet_2019}. Its clinical relevance is well-established, with reduced P3b amplitude and prolonged latency reliably distinguishing healthy individuals from those with Mild Cognitive Impairment (MCI) and Alzheimer's Disease (AD), making it a critical benchmark for assessing source localization accuracy \cite{Hedges_2016, Morrison_2019}. 
	
	Current pipelines often employ whole-brain analyses that aggregate signals across all regions. Such indiscriminate pooling conflates task-specific P300 generators with task-irrelevant noise, degrading separability and introducing systematic error. Achieving robust single-trial classification accuracy therefore requires spatially selective strategies that isolate neurophysiologically relevant activity while rejecting non-target contributions. This challenge is reflected in the wide accuracy range (53.54\% -- 84.00\%) reported across methodologies. A mini-review of the state-of-the art shows mean classification accuracies of 73.31\% with a confidence interval (CI) of $\pm$ 17.39\% (Student-t corrected, k = 5 studies, 95\% confidence level) \cite{Cecotti_2010, Syan_2010, Leoni_2022, Du_2023, Sharma_2017}.

	While most BCIs prioritize electrode reduction for usability, this approach often fails to capture the spatially distributed nature of P300 generators, necessitating broader sensor coverage for accurate neurophysiological measurement \cite{McCann_2015, Song_2015}. Moreover, both biological (e.g., eye movements) and environmental artifacts degrade signal quality - a critical limitation for single-trial analyses where SNR is paramount. Despite the P300's importance in cognitive research, inconsistent preprocessing and validation methodologies continue to impede reproducibility and hinder clinical translation.
	
	Accurate identification of these clinical biomarkers requires advanced source localization. eLORETA offers theoretically perfect localization under ideal conditions through its adaptive weighting approach, which suppresses noisy components while maintaining spatial precision across cortical depths. This capability proves particularly valuable for distinguishing deep and adjacent neural sources -- a critical requirement for anatomically precise applications \cite{PascualMarqui_2018}.

	
	Region-of-interest (ROI) selection addresses two fundamental limitations in EEG-based BCIs: (1) contamination from non-task-relevant neural activity and (2) inter- and intra-subject variability stemming from spatially diffuse signal sources. By prioritizing eLORETA-derived ROIs, this work's goal is to minimize noise inherent to whole-brain approaches while isolating neurophysiologically meaningful P300 activity. A simplified classification strategy -- based on peak amplitude timing within ROIs -- serves to directly evaluate the impact of spatial selectivity, deliberately avoiding algorithmic complexity that could obscure interpretability.

	\begin{figure}[thpb]
		\centering
		\includegraphics[width=8.5cm]{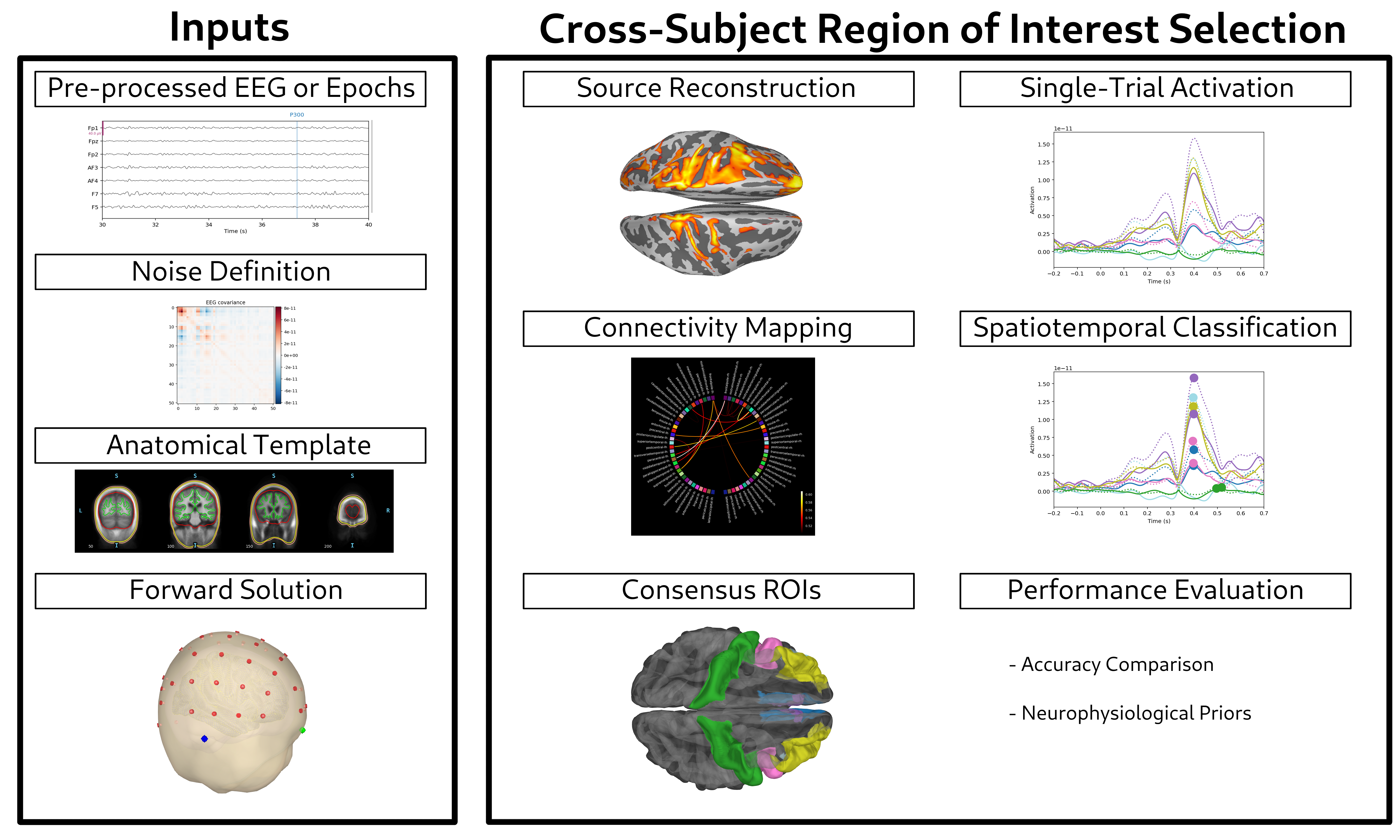}
		\caption{Consensus-Driven Hybrid Framework for Single-Trial P300 Detection Based on Cross-Subject Region of Interest (ROI) Selection}
		\label{fig:Flowchart}
	\end{figure}
	
	The current investigation aims to systematically evaluate the efficacy of ROI selection in improving single-trial P300 classification accuracy. First, we hypothesize that the classification performance reflects the precision of source localization. Second, we identify stable ROIs through eLORETA-based functional connectivity, testing their consistency across subjects and sessions to ensure generalizability. Finally, we compare discriminative power between ROI-driven and whole-brain approaches, positing that spatially focused analysis reduces contamination from noise-prone regions. Figure \ref{fig:Flowchart} illustrates the consensus-driven hybrid framework employed in this study, highlighting the integration of cross-subject ROI selection and spatiotemporal classification. By anchoring these objectives in the P300’s properties, the study aims at developing a reproducible framework for robust target-vs-non-target classification, adaptable to downstream clinical or consumer applications.
	
	\section{METHODS}
	
	\subsection{Study Design}
	The Eye-BCI multimodal dataset \cite{eye_bci_multi_dataset} includes 62-channel EEG recordings (10-10 system) from 31 healthy participants (age range: 22--57 years; 29 right-handed) during 2,520 P300 trials, enabling precise eLORETA source analysis for ROI validation. The dataset features synchronized EOG/EMG/eye-tracking \cite{GuttmannFlury_Dataset_2025} and was powered using Fitted Distribution Monte Carlo (FDMC) for sample size estimation \cite{GuttmannFlury_APriori_2019}. All experimental protocols are openly available \href{https://github.com/QinXinlan/EEG-experiment-to-understand-differences-in-blinking/}{https://github.com/QinXinlan/EEG-experiment-to-understand-differences-in-blinking/}, ensuring full reproducibility.

	\subsection{Signal Preprocessing Pipeline}
	
	The preprocessing pipeline optimizes data quality  through sequential artifact correction, spatial re-referencing, and temporal filtering. Initial artifact mitigation employs the Adaptive Blink Correction and De-drifting (ABCD) algorithm \cite{GuttmannFlury_ABC_2019} to address blink contamination and low-frequency drift. Channels demonstrating abnormal blink attenuation or bridging artifacts are excluded, systematically removing 11 malfunctioning electrodes across all recordings \cite{GuttmannFlury_BadChannel_2025}.

	Following artifact removal, low-frequency drifts ($<$ 0.1 Hz) induced by drying conductive gel or skin potentials are corrected via cubic spline interpolation, preserving millisecond-scale temporal fidelity of the P300 waveform. Signals are then re-referenced using a surface Laplacian filter to enhance spatial contrast by attenuating volume-conducted noise and emphasizing local cortical generators. Temporal filtering isolates neurophysiologically relevant spectral content: a zero-phase 4th-order Butterworth bandpass decomposes the signal into canonical frequency bands (Delta: 0.5 -- 4 Hz, Theta: 4 -- 7.5 Hz, Alpha: 7.5 -- 12.5 Hz, Beta: 12.5 -- 30 Hz, and P300’s characteristic range: 1 -- 15 Hz). Data that bypass this stage are marked “Unfiltered” in Table \ref{tab:classif}. The pre-processing pipeline concludes with epoch segmentation ($-\,200$ to $+\,700$ ms relative to stimulus onset), retaining pre-stimulus baselines for normalization.
	
	\subsection{Source Localization and Anatomical Modeling}
	Source localization transforms scalp EEG into cortical generator estimates by solving the ill-posed inverse problem through biophysical modeling and regularization. Unlike sensor-space analysis that struggles with overlapping sources, this approach reconstructs latent neural dynamics by incorporating anatomical constraints and electromagnetic propagation principles. Distributed methods like eLORETA provide whole-brain voxel-wise reconstructions without a priori assumptions, trading some spatial specificity for comprehensive coverage. The FreeSurfer fsaverage template serves as the anatomical reference, with cortical surfaces aligned through spherical morphing to preserve sulcal/gyral topology while minimizing cross-subject distortion \cite{Fischl_1999}. 
	
	The forward solution is generated using MNE-Python's boundary element model (BEM), incorporating tissue-specific electrical conductivity values (scalp: $0.33\,S/m$, skull: $0.006\,S/m$, brain: $0.33\,S/m$) to model EEG signal propagation through head tissues (Figure \ref{fig:Flowchart}). A discretized source space of 10,242 dipoles per hemisphere (3.1 mm spacing) balances spatial resolution for P300 generator discrimination with computational efficiency. This population-based approach reduces anatomical variability while maintaining precision for reliable region-of-interest identification \cite{Gramfort_2013}.

	\subsection{Functional Connectivity Analysis}

	Functional connectivity analysis reveals the transient network dynamics governing P300 generation through phase synchronization measures. Unlike amplitude-based correlations, phase-locking metrics effectively mitigate artifacts from volume conduction, making them ideal for single-trial EEG analysis. The approach leverages FreeSurfer's 68-region cortical parcellation, with time-locked extraction of source signals during P300 subcomponents (P3a: 150 -- 250ms; P3b: 300 -- 500ms) to isolate distinct network interactions.

	Three phase-locking metrics are evaluated for single-trial robustness: (1) the Phase Lag Index (PLI), capturing consistent timing differences while ignoring instantaneous synchrony \cite{Stam_2007}; (2) the weighted PLI (wPLI), emphasizing reliable phase differences to reduce noise sensitivity \cite{Vinck_2011}; and (3) the directed PLI (dPLI), quantifying directional influences in hierarchical interactions \cite{Stam_2012} (Table \ref{tab:metrics}). Each metric demonstrates unique advantages in directionality, noise robustness, and resistance to volume conduction. Following the analytical framework (Figure \ref{fig:Flowchart}), stable regions of interest were identified as the three brain regions most consistently observed among the top 20 connection pairs across sessions.
	
	\begin{table}[htbp]
		\centering
		\caption{Comparative Properties of Phase-Based Connectivity Metrics}
		\label{tab:metrics}
		\begin{tabular}{ccccc}
			\toprule
			\textbf{Metric}  & \textbf{\makecell{Direction-\\ality}} & \textbf{\makecell{Noise\\ Robustness}} & \textbf{\makecell{Volume \\ Conduction\\ Resistance}} & \textbf{\makecell{Recommen-\\ded Use \\Case}} \\
			\midrule
			PLI    & No    & Moderate & High & \makecell{Static\\ connectivity} \\
			\\
			wPLI   & No    & High     & High & \makecell{High-\\noise \\ environments} \\
			\\
			dPLI   & Yes   & Moderate & High & \makecell{Directed \\network \\analysis} \\
			\bottomrule
		\end{tabular}
	\end{table}

	Single-trial P300 classification is performed by identifying activation maxima within cross-subject ROIs across the entire epoch ($-\,200$ to $+\,700$ ms). Trials are classified as P300-positive if their global maxima fall within the neurophysiologically plausible window ($+\,200$ to $+\,550$ ms; labeled Count in Table \ref{tab:classif}). To enhance reliability, two additional criteria are applied: (1) a temporal agreement criterion requiring maxima clustering within the target window across most ROIs (Max), and (2) a hybrid approach combining both spatial and temporal criteria (Hybrid). This multi-stage validation leverages eLORETA's anatomical precision while accounting for individual session variability.
	
	This dual-layered approach addresses two critical challenges in single-trial analysis: (1) spatial variability in ROI relevance across sessions, mitigated by cross-session consistency in hub selection, and (2) temporal jitter in single-trial P300 latencies, countered by consensus-based timing validation. This method's innovation lies in its synergistic integration of spatially stable ROIs with dynamic temporal agreement, effectively reducing session-specific noise -- a significant improvement over conventional frameworks.

	\section{RESULTS}
	
	\subsection{Accuracy Comparison}
	
	Table \ref{tab:classif} summarizes classification accuracy across frequency bands and connectivity metrics, revealing systematic performance variations rooted in the P300’s neurophysiological composition and methodological properties of phase-based analyses. The P300 potential is generated through phase-locked delta and theta-range synchronized oscillations, which reflect its biphasic structure: the early P3a (frontal attention)  and later P3b (parietal memory updating). These mechanisms are further modulated by task-induced alpha activity, which gates attentional resources and consolidates contextual memory traces \cite{Berti_2016, Verleger_2020}.

	\begin{table}[htbp]
		\centering
		\caption{Classification Performance Across Frequency Bands and Methods For All 63 Sessions: Accuracy and Confidence Interval in \% (with 95\% Confidence Level) }
		\label{tab:classif}
		\begin{tabular}{lcccccc}
			\toprule
			\textbf{Method} & \textbf{P300} & \textbf{Delta} & \textbf{Theta} & \textbf{Alpha} & \textbf{Beta} & \textbf{\makecell{Unfilt-\\ered}} \\
			\midrule
			\multicolumn{7}{l}{\textit{Whole-brain}} \\
			\cmidrule(lr){1-7}
			Count & \makecell{\textbf{72.16}\\ \textbf{$\pm$ 2.09}}  & \makecell{62.79\\$\pm$ 2.01 } & \makecell{70.33\\$\pm$ 1.96 } & \makecell{66.99\\$\pm$ 1.63 } & \makecell{68.27\\$\pm$ 1.60 } & \makecell{65.32\\$\pm$ 2.07 } \\
			\\
			Max & \makecell{\textbf{70.13}\\\textbf{$\pm$ 1.96} } & \makecell{62.20\\$\pm$ 1.98 } & \makecell{67.47\\$\pm$ 1.81 } & \makecell{63.94\\$\pm$ 1.61 } & \makecell{64.37\\$\pm$ 1.89 } & \makecell{63.42\\$\pm$ 1.93 } \\
			\\
			Hybrid & \makecell{\textbf{77.36}\\\textbf{$\pm$ 1.77} } & \makecell{69.11\\$\pm$ 1.99 } & \makecell{76.84\\$\pm$ 1.81 } & \makecell{73.03\\$\pm$ 1.628 } & \makecell{74.06\\$\pm$ 1.59 } & \makecell{70.57\\$\pm$ 1.90 } \\
			
			\midrule
			
			\multicolumn{7}{l}{\textit{Phase Lag Index (PLI)}} \\
			\cmidrule(lr){1-7}
			Count & \makecell{\textbf{75.97}\\\textbf{$\pm$ 1.73} } & \makecell{65.48\\$\pm$ 2.14 } & \makecell{71.92\\$\pm$ 1.79 } & \makecell{68.62\\$\pm$ 1.68 } & \makecell{70.29\\$\pm$ 1.77 } & \makecell{69.77\\$\pm$ 2.15 } \\
			\\
			Max & \makecell{\textbf{69.93}\\\textbf{$\pm$ 2.14} } & \makecell{60.84\\$\pm$ 1.68 } & \makecell{67.99\\$\pm$ 1.60 } & \makecell{64.93\\$\pm$ 1.72 } & \makecell{65.84\\$\pm$ 1.57 } & \makecell{61.98\\$\pm$ 1.86 } \\
			\\
			Hybrid & \makecell{\textbf{81.57}\\\textbf{$\pm$ 1.65} } & \makecell{71.68\\$\pm$ 1.84 } & \makecell{77.36\\$\pm$ 1.56 } & \makecell{74.69\\$\pm$ 1.61 } & \makecell{76.01\\$\pm$ 1.61 } & \makecell{74.93\\$\pm$ 2.03 } \\
			
			\midrule
			
			\multicolumn{7}{l}{\textit{Weighted Phase Lag Index (wPLI)}} \\
			\cmidrule(lr){1-7}
			Count & \makecell{71.48\\$\pm$ 2.09 } & \makecell{62.35\\$\pm$ 2.23 } & \makecell{\textbf{72.55}\\\textbf{$\pm$ 1.85} } & \makecell{70.37\\$\pm$ 1.69 } & \makecell{71.96\\$\pm$ 1.67 } & \makecell{64.92\\$\pm$ 2.00 } \\
			\\
			Max & \makecell{\textbf{68.58}\\\textbf{$\pm$ 2.20} } & \makecell{62.11\\$\pm$ 2.03 } & \makecell{66.56\\$\pm$ 1.81 } & \makecell{65.60\\$\pm$ 1.73 } & \makecell{65.56\\$\pm$ 1.70 } & \makecell{62.58\\$\pm$ 2.04 } \\
			\\
			Hybrid  & \makecell{74.38\\$\pm$ 2.03 } & \makecell{66.40\\$\pm$ 2.17 } & \makecell{78.35\\$\pm$ 1.90 } & \makecell{77.04\\$\pm$ 1.50 } & \makecell{\textbf{79.46}\\\textbf{$\pm$ 1.62} } & \makecell{68.10\\$\pm$ 1.78 } \\
			
			\midrule
			
			\multicolumn{7}{l}{\textit{Directed Phase Lag Index (dPLI)}} \\
			\cmidrule(lr){1-7}
			Count & \makecell{\textbf{73.86}\\\textbf{$\pm$ 1.79} } & \makecell{64.81\\$\pm$ 1.95 } & \makecell{72.16\\$\pm$ 1.62 } & \makecell{71.29\\$\pm$ 1.70 } & \makecell{68.23\\$\pm$ 1.82 } & \makecell{68.26\\$\pm$ 2.04 } \\
			\\
			Max & \makecell{\textbf{69.73}\\\textbf{$\pm$ 1.96} } & \makecell{61.91\\$\pm$ 2.05 } & \makecell{65.25\\$\pm$ 1.98 } & \makecell{64.38\\$\pm$ 1.80 } & \makecell{64.78\\$\pm$ 1.73 } & \makecell{63.02\\$\pm$ 2.02 } \\
			\\
			Hybrid & \makecell{\textbf{79.90}\\\textbf{$\pm$ 1.83} } & \makecell{71.68\\$\pm$ 1.86 } & \makecell{77.91\\$\pm$ 1.62 } & \makecell{78.24\\$\pm$ 1.75 } & \makecell{74.23\\$\pm$ 1.70 } & \makecell{73.94\\$\pm$ 1.84 } \\
			
			\bottomrule
		\end{tabular}
	\end{table}

	PLI and dPLI excel in isolating these canonical P300 dynamics by prioritizing non-zero phase-lagged interactions. Delta and theta oscillations exhibit subtle temporal offsets (e.g., frontal theta preceding parietal alpha) that PLI detects by excluding volume-conduction-induced zero-lag correlations -- artifacts arising when signals from a single neural source propagate instantaneously to multiple electrodes. dPLI extends this by resolving directional dependencies (e.g., frontal → parietal), achieving 81.57\% (CI: $\pm$ 1.65\%) and 79.90\% (CI: $\pm$ 1.83\%) accuracy in the 1 -- 15 Hz band. 
	
	wPLI, in contrast, employs magnitude-weighted phase differences, prioritizing sustained oscillations over transient phase-locked activity. While this reduces its specificity to the P300’s core mechanisms (74.38\% $\pm$ 8.23\% accuracy in 1--15 Hz), it enhances robustness to noise in theta (72.55\% $\pm$ 7.5\%) and beta (79.46\% $\pm$ 6.55\%) bands. Theta oscillations, critical for continuous attentional maintenance, and beta activity, associated with task-set stability, involve less precise phase-lagged relationships, favoring wPLI’s smoothing of phase transitions \cite{Kardos_2014}.

	\subsection{Neurophysiological Priors}
		
\begin{figure}[thpb]
		\centering
		\includegraphics[width=8.2cm]{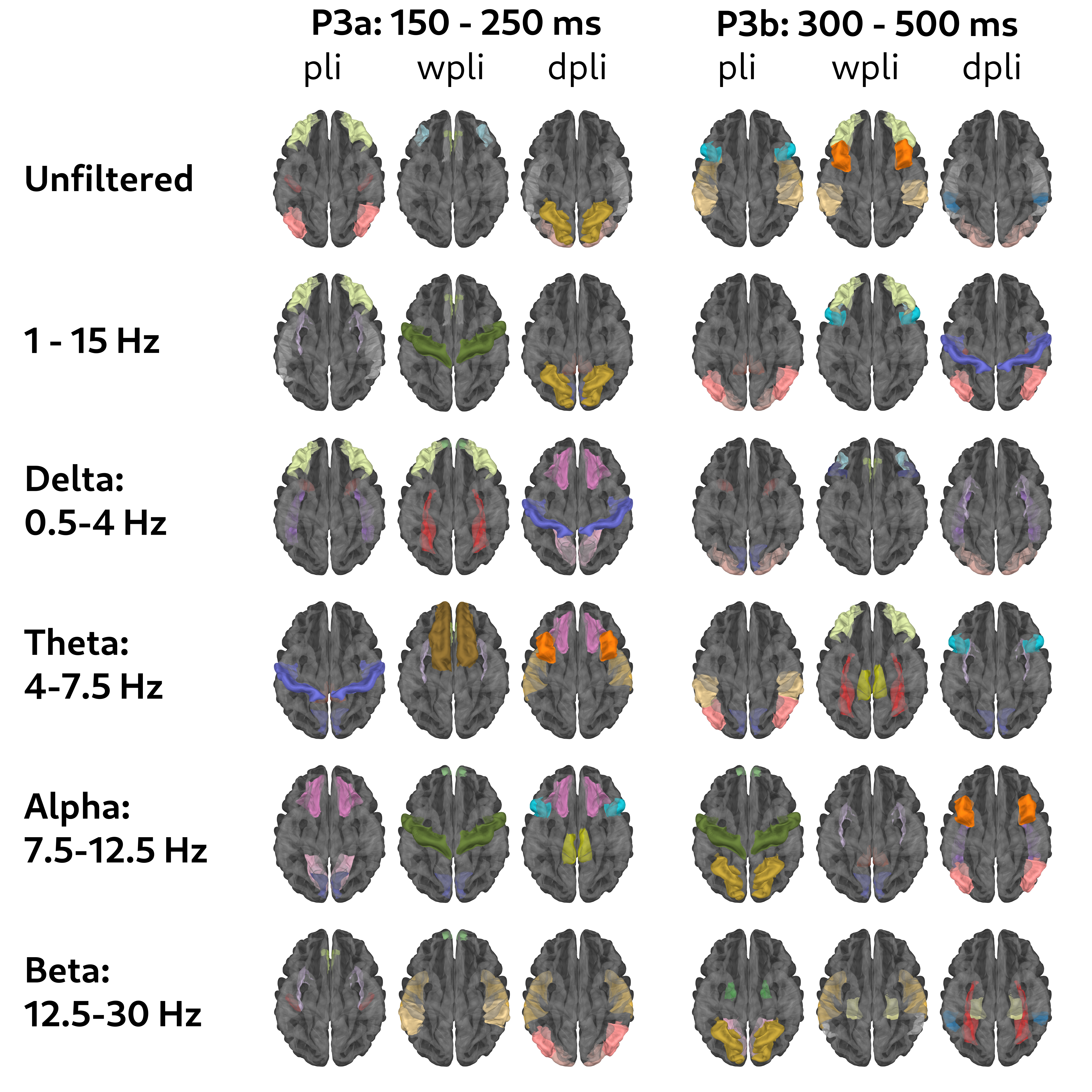}
		\caption{Spatiotemporal Dynamics of P300: Cross-Subject Functional Hubs via Phase Synchronization }
		\label{fig:Priors}
	\end{figure}

	Connectivity patterns mirror the hierarchical cascade underlying the visual P300. Phase-lagged delta -- theta coupling binds rostral middle frontal and anterior insular sources to initiate attentional capture (P3a), whereas alpha -- theta coherence between middle temporal and inferior parietal cortices sustains context-updating operations (P3b). As depicted in Figure \ref{fig:Priors} these transient, non-zero-lag interactions gate distractors and stabilize target representations \cite{Citherlet_2019}.
	
	dPLI reveals a posterior executive hierarchy where top-down modulation from the isthmus cingulate regulates parietal attentional resources, while bottom-up signaling from primary visual areas propagates retinotopic salience for rapid target discrimination \cite{Jiang_2023}. Complementarily, wPLI identifies beta-band coherence in superior temporal, supramarginal, and frontal pole regions: frontal oscillations sustain attentional focus, while supramarginal synchronization stabilizes multisensory integration and predictive templates for efficient target prioritization \cite{Proskovec_2018, Rossi_2023}.

	Collectively, these connectivity patterns delineate the P300’s spatiotemporal architecture: theta/alpha networks mediate transient attention-memory interactions, while beta oscillations sustain cognitive frameworks for target detection. This hierarchical integration of frequency-specific dynamics validates the P300 as a marker of distributed neural coordination, bridging perceptual salience with memory-driven prioritization in visual paradigms.

	\section{DISCUSSION}
	
	\subsection{Neurophysiological Insights}
	
	The P300’s spatiotemporal signature embodies cortical decision-making: an early frontal delta -- theta burst flags stimulus salience (P3a), followed by parietal alpha -- theta coupling that embeds the event in working memory (P3b). This sequence respects the canonical cortical hierarchy -- sensory input is first tagged for relevance, then integrated with context \cite{Bachman_2018} -- and is reinforced by intracranial data showing that the response emerges from precisely timed, cross-regional interactions rather than local amplitude jumps, with concurrent beta activity in superior temporal and frontal sites sustaining attention \cite{Krusienski_2011}.
		
	The observed frontoinsular-parietal phase-lagged connectivity suggests the P300 functions as a cortical reset mechanism, mediated by theta-driven long-range synchronization in key hubs (insula, cingulate). Connectivity metrics reveal distinct processing layers: PLI/dPLI track transient theta -- alpha interactions (core P300 signatures), while wPLI detects stabilizing beta-band activity. This frequency-specific specialization reflects the P300's neurophysiological architecture -- transient stimulus processing embedded within sustained cognitive control networks.

	\subsection{Clinical Translation}
	The P300's robust spatiotemporal properties and sensitivity to distributed network disruptions make it a compelling candidate for neurodegenerative disease biomarkers. Our findings demonstrate that source-localized ROI analysis precisely isolates disease-relevant neural hubs (e.g., insula and parietal regions) while effectively mitigating noise from non-task-related activity. This approach provides a refined method for detecting MCI and predicting conversion to AD by clearly distinguishing pathological network disruptions from normal variability. Clinical validation studies have further strengthened this application, showing that P300 latency prolongation and amplitude reduction correlate strongly with early AD pathology \cite{Jervis_2019}.
	
	Equally important, the P300's inherent insensitivity to educational and cultural biases makes it particularly valuable for diverse clinical populations. While its diagnostic utility for Alzheimer's disease is well-established, the component's rich spatiotemporal dynamics reveal broader clinical potential. The observed beta-band coherence in frontal-parietal networks, for instance, may enable novel applications ranging from localizing epileptic network disruptions to tracking cognitive decline in Parkinson's disease. This combination of diagnostic reliability and methodological versatility positions the P300 as a uniquely adaptable tool for both research and clinical neurology \cite{Parra_2012}.

	\subsection{Limitations and Future Directions}
	
	Our framework achieves an explainable peak classification accuracy of 81.57\% (CI $\pm$ 1.65\%) in the 1 -- 15 Hz band, surpassing the current state-of-the-art mean (73.31\% $\pm$ 17.39\%). This improvement stems from two key innovations: (1) eLORETA-derived ROI selection isolating phase-lagged interactions in functional hubs (e.g., insula, parietal), and (2) hybrid spatiotemporal validation combining peak detection with cross-regional consensus. Our anatomically constrained method directly ties classification decisions to activations in neurophysiologically relevant networks, providing both higher accuracy and mechanistic interpretability. 
	
	The present investigation offers valuable insights into cross-subject consensus approaches for P300 analysis, while naturally paving the way to explore inter- and intra-subject variability as both a source of complementary diagnostic information and potential noise. The framework developed in this work provides a solid foundation for such exploration -- subsequent research could build upon our protocol while incorporating individualized connectivity patterns and dynamic network analyses to reveal clinically relevant biomarkers within subject-specific variations. The fsaverage template's standardization, which proved effective for establishing methodological validity, now creates opportunities to examine how anatomical personalization might enhance precision, especially in clinical populations where tissue characteristics may differ from healthy baselines.
	
	The consistent performance of our heuristic ROI selection demonstrates the robustness of the approach while inviting systematic optimization. Future work could transform our current parameters into adaptive frameworks where ROI cardinality and temporal windows dynamically adjust based on individual connectivity profiles. This methodological evolution, combined with multimodal validation against established biomarkers and longitudinal tracking of disease progression, promises to translate these findings obtained from healthy participants into potentially clinically actionable tools. The current study’s  classification accuracy with standardized parameters suggests that future refinements will yield even greater diagnostic potential while maintaining the approach’s practical accessibility.

	\section{CONCLUSIONS}
	
	This study demonstrates that targeted region-of-interest (ROI) analysis, grounded in eLORETA-based source localization and cross-subject functional connectivity, consistently improves single-trial P300 classification accuracy over conventional whole-brain methods. The framework achieves robust performance (81.57\% $\pm$ 6.69\% in 1--15 Hz range) by isolating task-relevant hubs such as the insula and parietal regions, validating the hypothesis that spatial selectivity enhances biomarker detection. Its hybrid spatiotemporal criterion balances anatomical specificity with temporal consensus, addressing variability inherent to single-trial EEG while preserving millisecond-level precision.
	
	The work’s originality lies in synthesizing population-level anatomical constraints, phase-resolved connectivity metrics, and adaptive classification to resolve the spatial-noise trade-off in non-invasive BCIs. This approach bridges cognitive neuroscience with clinical translation by mapping the P300’s hierarchical network dynamics. Its insensitivity to cultural/educational biases and compatibility with portable systems underscore its potential for global deployment in neurodegenerative disease diagnostics. Future integration of individualized anatomical refinements and multimodal biomarkers promises to extend these findings into precision tools for early intervention, marking a pivotal step toward clinically actionable neurophysiological monitoring.

	\addtolength{\textheight}{-12cm}   
	


	
	
	

	

\end{document}